\documentclass[letterpaper,conference,twocolumn]{IEEEtranBB}
\usepackage{amsmath,amssymb,mathrsfs}
\usepackage{graphicx}
\usepackage{psfrag}
\usepackage{cite}
\usepackage{multirow}
\usepackage{flushend}

\usepackage{calc}
\usepackage{graphicx}
        \graphicspath{{figs/}{matlab/}}   
\usepackage[usenames,dvipsnames]{color}

\usepackage[tight,small,bf]{subfigure}   

\usepackage{float}
\usepackage{caption}
\DeclareCaptionLabelSeparator{periodquad}{. \quad}
\usepackage{amsthm}
\usepackage[colorlinks=true,linkcolor=black,anchorcolor=black,citecolor=black,filecolor=black,menucolor=black,urlcolor=black]{hyperref}  

\newcommand{\anneq}[1]{\overset{\text{(#1)}}{=}}  
\newcommand{\annleq}[1]{\overset{\text{\!\!\!(#1)\!\!\!}}{\leq}}  
\newcommand{\inpmfspace}{\hspace{0.35ex}}

\usepackage{xspace}
\usepackage{bbm}
\catcode`@=11
\chardef\mathlig@atcode\count255

\def\actively#1#2{\begingroup\uccode`\~=`#2\relax\uppercase{\endgroup#1~}}
\def\mathlig@gobble{\afterassignment\mathlig@next@cmd\let\mathlig@next= }
\def\mathlig@delim{\mathlig@delim}
\def\mathlig@defcs#1{\expandafter\def\csname#1\endcsname}
\def\mathlig@let@cs#1#2{\expandafter\let\expandafter#1\csname#2\endcsname}
\def\mathlig@appendcs#1#2{\expandafter\edef\csname#1\endcsname{\csname#1\endcsname#2}}

\def\mathlig#1#2{\mathlig@checklig#1\mathlig@end\mathlig@defcs{mathlig@back@#1}{#2}\ignorespaces}

\def\mathlig@checklig#1#2\mathlig@end{%
 \expandafter\ifx\csname mathlig@forw@#1\endcsname\relax
 \expandafter\mathchardef\csname mathlig@back@#1\endcsname=\mathcode`#1%
 \mathcode`#1"8000\actively\def#1{\csname mathlig@look@#1\endcsname}%
 \mathlig@dolig#1\mathlig@delim
\fi
\mathlig@checksuffix#1#2\mathlig@end
}

\def\mathlig@checksuffix#1#2\mathlig@end{%
\ifx\mathlig@delim#2\mathlig@delim\relax\else\mathlig@checksuffix@{#1}#2\mathlig@end\fi
}
\def\mathlig@checksuffix@#1#2#3\mathlig@end{%
\expandafter\ifx\csname mathlig@forw@#1#2\endcsname\relax\mathlig@dosuffix{#1}{#2}\fi
\mathlig@checksuffix{#1#2}#3\mathlig@end
}

\def\mathlig@dosuffix#1#2{%
\mathlig@appendcs{mathlig@toks@#1}{#2}%
\mathlig@dolig{#1}{#2}\mathlig@delim
}

\def\mathlig@dolig#1#2\mathlig@delim{%
 \mathlig@defcs{mathlig@look@#1#2}{%
 \mathlig@let@cs\mathlig@next{mathlig@forw@#1#2}\futurelet\mathlig@next@tok\mathlig@next}%
 \mathlig@defcs{mathlig@forw@#1#2}{%
  \mathlig@let@cs\mathlig@next{mathlig@back@#1#2}%
  \mathlig@let@cs\checker{mathlig@chck@#1#2}%
  \mathlig@let@cs\mathligtoks{mathlig@toks@#1#2}%
  \expandafter\ifx\expandafter\mathlig@delim\mathligtoks\mathlig@delim\relax\else
  \expandafter\checker\mathligtoks\mathlig@delim\fi
  \mathlig@next
 }%
 \mathlig@defcs{mathlig@toks@#1#2}{}%
 \mathlig@defcs{mathlig@chck@#1#2}##1##2\mathlig@delim{%
  \ifx\mathlig@next@tok##1%
   \mathlig@let@cs\mathlig@next@cmd{mathlig@look@#1#2##1}\let\mathlig@next\mathlig@gobble
  \fi 
  \ifx\mathlig@delim##2\mathlig@delim\relax\else
   \csname mathlig@chck@#1#2\endcsname##2\mathlig@delim
  \fi
 }%
 \ifx\mathlig@delim#2\mathlig@delim\else
  \mathlig@defcs{mathlig@back@#1#2}{\csname mathlig@back@#1\endcsname #2}%
 \fi
}%

\catcode`@\mathlig@atcode

\newcommand{\muspace}{\mspace{1mu}}

\DeclareRobustCommand{\scond}{\mathchoice{\muspace\vert\muspace}{\vert}{\vert}{\vert}}
\mathlig{|}{\scond}
\newcommand{\cond}{\mathchoice{\,\vert\,}{\mspace{2mu}\vert\mspace{2mu}}{\vert}{\vert}}

\DeclareRobustCommand{\discint}{\mathchoice{\mspace{-1.5mu}:\mspace{-1.5mu}}{\mspace{-1.5mu}:\mspace{-1.5mu}}{:}{:}}
\mathlig{::}{\discint}

\newcommand{\Cc}{\mathcal{C}}

\newcommand{\Pc}{\mathcal{P}}

\newcommand{\Xc}{\mathcal{X}}

\newcommand{\Rr}{\mathscr{R}}

\newcommand{\pen}{{P_e^{(n)}}}

\newcommand{\aep}{{\mathcal{T}_{\epsilon}^{(n)}}}

\newcommand{\Mh}{{\hat{M}}}

\newcommand{\mh}{{\hat{m}}}

\def\e{\epsilon}
\def\eps{\epsilon}

\let\P\relax
\DeclareMathOperator\P{\textsf{P}}

\def\textiid{i.i.d.\@\xspace}
\newcommand\iid{\ifmmode\text{ i.i.d. } \else \textiid \fi}

\def\mathllap{\mathpalette\mathllapinternal}
\def\mathllapinternal#1#2{%
  \llap{$\mathsurround=0pt#1{#2}$}}

\def\clap#1{\hbox to 0pt{\hss#1\hss}}
\def\mathclap{\mathpalette\mathclapinternal}
\def\mathclapinternal#1#2{%
  \clap{$\mathsurround=0pt#1{#2}$}}

\let\oldstackrel\stackrel
\renewcommand{\stackrel}[2]{\oldstackrel{\mathclap{#1}}{#2}}

\renewcommand{\hbar}{h\mathllap{\overline{\vphantom{h}\hphantom{\rule{4.6pt}{0pt}}}\mspace{0.77mu}}}

\catcode`~=11 
\newcommand{\urltilde}{\kern -.06em\lower -.06em\hbox{~}\kern .02em}
\catcode`~=13 

\theoremstyle{definition}  
\newtheorem{theor}{Theorem}

\title{A Comparison of Superposition Coding Schemes}
\author{
\authorblockN{Lele Wang,  Eren \c{S}a\c{s}o\u{g}lu, 
Bernd Bandemer, and Young-Han Kim
}
\authorblockA{Department of Electrical and Computer Engineering\\
University of California, San Diego\\
La Jolla, CA 92093, USA\\
Email: \{lew001, esasoglu, bandemer, yhk\}@ucsd.edu}}

\hypersetup{
	pdfauthor = {Lele Wang, Eren Sasoglu, Bernd Bandemer, and Young-Han Kim},
	pdftitle = {A Comparison of Two Superposition Coding Schemes},
	pdfsubject = {Information theory},
	pdfkeywords = {},
	pdfcreator = {LaTeX}
}

\begin{document}
\maketitle

\begin{abstract}
There are two variants of superposition coding schemes. Cover's original superposition
coding scheme has code clouds of the identical shape, while Bergmans's superposition coding
scheme has code clouds of independently generated shapes. These two schemes
yield identical achievable rate regions in several scenarios, such as
the capacity region for degraded broadcast channels.
This paper shows that under the optimal maximum likelihood decoding, 
these two superposition coding schemes can result in different rate regions.
In particular, it is shown that for the two-receiver broadcast channel,
Cover's superposition coding scheme can achieve rates strictly larger than
Bergmans's scheme.
\end{abstract}

\vspace{3mm}
\section{Introduction}

Superposition coding is one of the fundamental building blocks of
coding schemes in network information theory. This idea was first
introduced by Cover in 1970 at the IEEE International Symposium on
Information Theory, Noordwijk, the Netherlands, in a talk titled
``Simultaneous Communication,'' and appeared in his 1972
paper~\cite{Cover1972}.  Subsequently, Bergmans~\cite{Bergmans1973}
adapted Cover's superposition coding scheme to the general degraded
broadcast channel (this scheme is actually applicable to any
nondegraded broadcast channel), which establishes the capacity region
along with the converse proof by Gallager~\cite{Gallager1974}.  Since
then, superposition coding has been applied in numerous problems,
including multiple access channels~\cite{Grant--Rimoldi--Urbanke--Whiting2001}, interference
channels~\cite{Carleial1978, Han--Kobayashi1981,
Chong--Motani--Garg--El-Gamal2008}, relay
channels~\cite{Cover--El-Gamal1979}, channels
with feedback~\cite{Cover--Leung1981, Ozarow1984}, and wiretap
channels~\cite{Csiszar--Korner1978, Chia--El-Gamal2009}.

%

In a nutshell, the objective of superposition coding is 
to communicate two message simultaneously by
encoding them into a single signal in two layers.
A ``better'' receiver of the signal can then recover the messages on both layers while a 
``worse'' receiver can recover the message on the coarse layer of the signal and
ignore the one on the fine layer. 

\begin{figure}[t]
	\subfigure[Homogeneous coding]{
		\hspace*{-1mm}
		\includegraphics{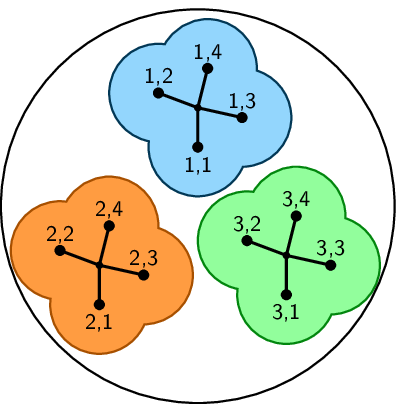}
		\label{fig:clouds_uv}
	} 
	\hfill
	\subfigure[Heterogeneous coding]{
		\includegraphics{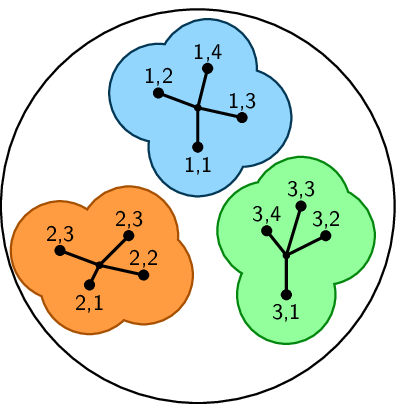}
		\hspace*{-2mm}
		\vspace*{2mm}
		\label{fig:clouds_ux}
	}
	\caption{Superposition codebooks for which (a) the structure within each cloud is identical and (b) the structure is nonidentical between clouds. Codewords (dots) are annotated by ``$m_1,m_2$'', where $m_1$ is the coarse layer message and $m_2$ is the fine layer message.}
	\label{fig:twoKindsOfSuperpositionCoding}
\end{figure}

On a closer look, there are two variants of the superposition coding idea 
in the literature, which differ in how the codebooks are generated. 
The first variant is described in Cover's original 1972 paper~\cite{Cover1972}. Both messages are first encoded independently via separate random codebooks of auxiliary sequences. To send a message pair, the auxiliary sequences associated with each message are then mapped through a symbol-by-symbol superposition function (such as addition) to generate the actual codeword.
One can visualize 
the image of one of the codebooks centered around a fixed codeword from the other 
as a ``cloud'' (see the illustration in Figure~\ref{fig:clouds_uv}).
Since all clouds are images of the same random codebook (around different cloud centers), 
we refer to this variant as \emph{homogeneous superposition coding}. 
Note that in this variant, both messages enter on an equal footing
and the corresponding auxiliary sequences play the same role.
Thus, there is no natural distinction between ``coarse'' and ``fine'' layers and there are two ways to group the resulting superposition codebook into clouds.

The second variant was introduced in Bergmans's 1973 paper~\cite{Bergmans1973}. 
Here, the coarse message is encoded in a random codebook of auxiliary sequences. For each auxiliary sequence, a random satellite codebook is generated conditionally independently to represent the fine layer message. This naturally results in clouds of codewords given each such satellite codebook. Since all clouds are generated independently, 
we refer to this variant as \emph{heterogeneous superposition coding}.
This is illustrated in Figure~\ref{fig:clouds_ux}.

A natural question is whether these two variants are fundamentally different, and if so, which of the two is preferable. 
Both variants achieve the capacity region of
the degraded broadcast channel~\cite{Bergmans1973}.
For the two-user-pair interference channel,
the two variants again achieve the identical Han--Kobayashi inner bound
(see \cite{Han--Kobayashi1981} for homogeneous superposition coding and
\cite{Chong--Motani--Garg--El-Gamal2008} for heterogeneous superposition coding).
Since heterogeneous superposition coding usually yields 
a simpler characterization of the achievable rate region with
fewer auxiliary random variables, it is tempting to prefer this variant.

In contrast, we show in this paper that homogeneous superposition
coding always achieves a rate region at least as large as that of
heterogeneous superposition coding for two-user broadcast channels,
provided that the optimal maximum likelihood decoding rule is used.
Furthermore, this dominance can be sometimes strict.  Intuitively
speaking, homogeneous superposition coding results in more structured
interference from the undesired layer, the effect of which
becomes tangible under optimal decoding. 

The rest of the paper is as follows. In Section~\ref{sec:superposition}, we formally define
the two variants of superposition coding schemes and present their respective rate regions.
In Section~\ref{sec:main}, we compare these rate regions. Additional remarks are provided in
Section~\ref{sec:remarks}.

Throughout the paper, we closely follow the notation in~\cite{El-Gamal--Kim2011}. 
In particular, for $X \sim p(x)$ and $\e \in (0,1)$, we define the set of $\e$-typical $n$-sequences
$x^n$ (or the typical set in short)~\cite{Orlitsky--Roche2001} as
$\aep(X) = \{ x^n : | \#\{i : x_i = x\}/n - p(x) | \le \e p(x)
\text{ for all } x \in \Xc \}$.

\vspace{3mm}
\section{Rate Regions for the Two-Receiver BC}
\label{sec:superposition} 
\begin{figure}[h]
\centering
\psfrag{m}[c]{\footnotesize $M_1,M_2$}
\psfrag{x}[b]{\footnotesize$X^n$}
\psfrag{p}[c]{\footnotesize$p(y_1,y_2|x)$}
\psfrag{y1}[b]{\footnotesize$Y_1^n$}
\psfrag{y2}[b]{\footnotesize$Y_2^n$}
\psfrag{m1}[b]{\footnotesize$\Mh_1$}
\psfrag{m2}[b]{\footnotesize$\Mh_2$}
\psfrag{e}[c]{\footnotesize Enc}
\psfrag{d1}[c]{\footnotesize Dec~1}
\psfrag{d2}[c]{\footnotesize Dec~2}
\includegraphics[scale=0.4]{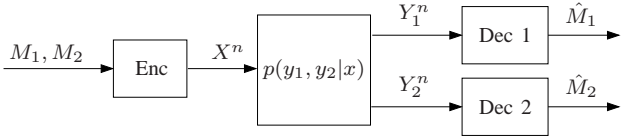}
\caption{Two-receiver broadcast channel.}
\label{fig:broadcast}
\end{figure}

Consider a two-receiver discrete memoryless broadcast channel
 depicted in Figure~\ref{fig:broadcast}.  The sender wishes to communicate
 message $M_1$ to receiver~1 and message $M_2$ to receiver~2. We define a $(2^{nR_1}, 2^{nR_2}, n)$ code by an encoder $x^n(m_1,m_2)$ and two receivers $\mh_1(y_1^n)$ and $\mh_2(y_2^n)$. We assume the message pair $(M_1,M_2)$ is uniform over $[1::2^{nR_1}]\times[1::2^{nR_2}]$ and independent of each other. The average probability of error is defined as
$\pen = \P\{(M_1, M_2)\neq (\hat{M}_1,\hat{M}_2)\}$. A rate pair $(R_1,R_2)$ is said to be achievable if
there exists a sequence of $(2^{nR_1},2^{nR_2},n)$ code such that $\lim_{n\to \infty}\pen = 0$.

We now describe the two superposition coding techniques for this
channel and compare their achievable rate regions under optimal
decoding.

\subsection{Homogeneous Superposition Coding ($UV$ Scheme)}
\emph{Codebook generation:} 
Fix a pmf $p(u)\inpmfspace p(v)$ and a function $x(u,v)$. Randomly and independently generate $2^{nR_1}$ sequences $u^n(m_1)$, $m_1 \in [1::2^{nR_1}]$, each according to $\prod_{i=1}^n p_U(u_i)$, and $2^{nR_2}$ sequences $v^n(m_2)$, $m_2 \in [1::2^{nR_2}]$, each according to $\prod_{i=1}^n p_V(v_i)$. 

\smallskip 
\emph{Encoding:} 
To send the message pair $(m_1,m_2)$, transmit $x_i(u_i(m_1),v_i(m_2))$ at time $i\in[1::n]$.

\smallskip 
\emph{Decoding:}
Both receivers use simultaneous nonunique decoding, which is rate-optimal in the sense that it achieves the same rate region as maximum likelihood decoding~\cite{Bandemer--El-Gamal--Kim2012a} under the codebook ensemble at hand.
In particular, upon receiving $y_1^n$, receiver~1 declares $\mh_1$ is sent if it is the unique message such that
$$
\big( u^n(\mh_1),v^n(m_2),y_1^n \big)
	\in\aep
$$
for some $m_2$. If there is no unique ~$\mh_1$, it declares an error.  Similarly, upon receiving $y_2^n$,
receiver~2 declares $\mh_2$ is sent if it is the unique message such that
$$
\big( u^n(m_1),v^n(\mh_2),y_2^n \big)
	\in\aep
$$
for some $m_1$.
If there is no unique~$\mh_2$, it declares an error.  Standard
typicality arguments show that receiver~1 will succeed if 
\begin{align}
\label{eqn:uv1}
R_1<I(U;Y_1)\quad\quad\text{or}\qquad
\begin{aligned}
R_1+R_2 &< I(X;Y_1) \\
R_1 &< I(X;Y_1\cond V),
\end{aligned}
\end{align}
or, equivalently, if
\begin{align*}
	R_1 &< I(X;Y_1 \cond V) \\
	R_1 + \min\{R_2,I(X;Y_1 \cond U)\} &< I(X;Y_1).
\end{align*}
Similarly, receiver~2 will succeed if
\begin{align}
\label{eqn:uv2}
R_2<I(V;Y_2)\quad\quad\text{or}\qquad
\begin{aligned}
R_1+R_2 &< I(X;Y_2) \\
R_2 &< I(X;Y_2\cond U),
\end{aligned}
\end{align}
or, equivalently, if
\begin{align*}
	R_2 &< I(X;Y_2 \cond U) \\
	R_2 + \min\{R_1,I(X;Y_2 \cond V)\} &< I(X;Y_2).
\end{align*}
The regions for both receivers are depicted in  Table~\ref{table:comparison}.  Letting $\Rr_{UV}(p)$ denote the set of rates $(R_1,R_2)$ satisfying
\eqref{eqn:uv1}~and~\eqref{eqn:uv2}, it follows that the rate
region 
$$
\Rr_{UV}=\text{co}\bigg(\bigcup_{p\in \Pc_{UV} }\Rr_{UV}(p)\bigg)
$$
is achievable.  Here, $\text{co}(\cdot)$ denotes
convex hull, and 
$\Pc_{UV}$ is the set of distributions of the
form $p=p(u)\inpmfspace p(v)\inpmfspace p(x|u,v)$ where $p(x|u,v)$ represents a deterministic function.

\begin{table*}[!t]
\normalsize
\centering
\begin{tabular}{ccc}
  \hline
  & &\\[-6pt]
  & Receiver~1 & Receiver~2  \\
  & & \\[-6pt]  
  \hline
  & &\\[-3pt]
  $\Rr_{UV}(p)$ & $\hspace{11.8em}R_1 < I(X;Y_1\cond V)$ & $\hspace{11.8em}R_2 < I(X;Y_2 \cond U)$\\
  $p=p(u)\inpmfspace p(v) \inpmfspace p(x|u,v)$ & $R_1 + \min\{R_2,I(X;Y_1 \cond U)\}< I(X;Y_1)$ &
  $R_2 + \min\{R_1,I(X;Y_2\cond V)\}< I(X;Y_2)$ \\
    & &\\[-6pt]
  &
  \def\svgscale{1.0}
\begingroup%
  \makeatletter%
  \providecommand\color[2][]{%
    \errmessage{(Inkscape) Color is used for the text in Inkscape, but the package 'color.sty' is not loaded}%
    \renewcommand\color[2][]{}%
  }%
  \providecommand\transparent[1]{%
    \errmessage{(Inkscape) Transparency is used (non-zero) for the text in Inkscape, but the package 'transparent.sty' is not loaded}%
    \renewcommand\transparent[1]{}%
  }%
  \providecommand\rotatebox[2]{#2}%
  \ifx\svgwidth\undefined%
    \setlength{\unitlength}{156.36406431bp}%
    \ifx\svgscale\undefined%
      \relax%
    \else%
      \setlength{\unitlength}{\unitlength * \real{\svgscale}}%
    \fi%
  \else%
    \setlength{\unitlength}{\svgwidth}%
  \fi%
  \global\let\svgwidth\undefined%
  \global\let\svgscale\undefined%
  \makeatother%
  \begin{picture}(1,0.5621168)%
    \put(0,0){\includegraphics[width=\unitlength]{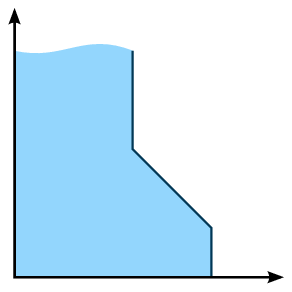}}%
    \put(0.78125838,0.01826174){\color[rgb]{0,0,0}\makebox(0,0)[lb]{\smash{$R_1$}}}%
    \put(0.25553193,0.52585978){\color[rgb]{0,0,0}\makebox(0,0)[rb]{\smash{$R_2$}}}%
  \end{picture}%
\endgroup%

  &
  \def\svgscale{1.0}
\begingroup%
  \makeatletter%
  \providecommand\color[2][]{%
    \errmessage{(Inkscape) Color is used for the text in Inkscape, but the package 'color.sty' is not loaded}%
    \renewcommand\color[2][]{}%
  }%
  \providecommand\transparent[1]{%
    \errmessage{(Inkscape) Transparency is used (non-zero) for the text in Inkscape, but the package 'transparent.sty' is not loaded}%
    \renewcommand\transparent[1]{}%
  }%
  \providecommand\rotatebox[2]{#2}%
  \ifx\svgwidth\undefined%
    \setlength{\unitlength}{156.36406351bp}%
    \ifx\svgscale\undefined%
      \relax%
    \else%
      \setlength{\unitlength}{\unitlength * \real{\svgscale}}%
    \fi%
  \else%
    \setlength{\unitlength}{\svgwidth}%
  \fi%
  \global\let\svgwidth\undefined%
  \global\let\svgscale\undefined%
  \makeatother%
  \begin{picture}(1,0.56004816)%
    \put(0,0){\includegraphics[width=\unitlength]{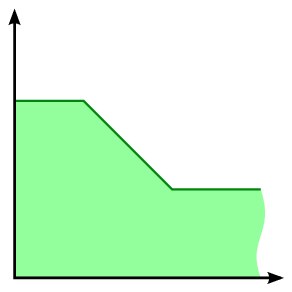}}%
    \put(0.78125837,0.01692635){\color[rgb]{0,0,0}\makebox(0,0)[lb]{\smash{$R_1$}}}%
    \put(0.25553192,0.52452439){\color[rgb]{0,0,0}\makebox(0,0)[rb]{\smash{$R_2$}}}%
  \end{picture}%
\endgroup%
\\  
  \hline
  & &\\[-3pt]
  $\Rr_{UX}(p)$ & $R_1+\min\{R_2, I(X;Y_1 \cond U)\} < I(X;Y_1)$& $\hspace{3.7em}R_2 < I(X;Y_2 \cond U)$  \\
  $p=p(u,x)$ & & $R_1+R_2 < I(X;Y_2)$\\
    & &\\[-6pt]
  &
  \def\svgscale{1.0}
\begingroup%
  \makeatletter%
  \providecommand\color[2][]{%
    \errmessage{(Inkscape) Color is used for the text in Inkscape, but the package 'color.sty' is not loaded}%
    \renewcommand\color[2][]{}%
  }%
  \providecommand\transparent[1]{%
    \errmessage{(Inkscape) Transparency is used (non-zero) for the text in Inkscape, but the package 'transparent.sty' is not loaded}%
    \renewcommand\transparent[1]{}%
  }%
  \providecommand\rotatebox[2]{#2}%
  \ifx\svgwidth\undefined%
    \setlength{\unitlength}{156.36406308bp}%
    \ifx\svgscale\undefined%
      \relax%
    \else%
      \setlength{\unitlength}{\unitlength * \real{\svgscale}}%
    \fi%
  \else%
    \setlength{\unitlength}{\svgwidth}%
  \fi%
  \global\let\svgwidth\undefined%
  \global\let\svgscale\undefined%
  \makeatother%
  \begin{picture}(1,0.56065047)%
    \put(0,0){\includegraphics[width=\unitlength]{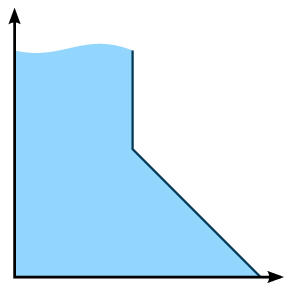}}%
    \put(0.78125833,0.01679548){\color[rgb]{0,0,0}\makebox(0,0)[lb]{\smash{$R_1$}}}%
    \put(0.25553188,0.52439353){\color[rgb]{0,0,0}\makebox(0,0)[rb]{\smash{$R_2$}}}%
  \end{picture}%
\endgroup%

  &
  \def\svgscale{1.0}
\begingroup%
  \makeatletter%
  \providecommand\color[2][]{%
    \errmessage{(Inkscape) Color is used for the text in Inkscape, but the package 'color.sty' is not loaded}%
    \renewcommand\color[2][]{}%
  }%
  \providecommand\transparent[1]{%
    \errmessage{(Inkscape) Transparency is used (non-zero) for the text in Inkscape, but the package 'transparent.sty' is not loaded}%
    \renewcommand\transparent[1]{}%
  }%
  \providecommand\rotatebox[2]{#2}%
  \ifx\svgwidth\undefined%
    \setlength{\unitlength}{156.36406439bp}%
    \ifx\svgscale\undefined%
      \relax%
    \else%
      \setlength{\unitlength}{\unitlength * \real{\svgscale}}%
    \fi%
  \else%
    \setlength{\unitlength}{\svgwidth}%
  \fi%
  \global\let\svgwidth\undefined%
  \global\let\svgscale\undefined%
  \makeatother%
  \begin{picture}(1,0.5613443)%
    \put(0,0){\includegraphics[width=\unitlength]{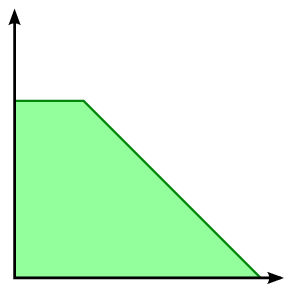}}%
    \put(0.78125837,0.01692633){\color[rgb]{0,0,0}\makebox(0,0)[lb]{\smash{$R_1$}}}%
    \put(0.25553192,0.52452438){\color[rgb]{0,0,0}\makebox(0,0)[rb]{\smash{$R_2$}}}%
  \end{picture}%
\endgroup%
\\  
  \hline
\end{tabular}
\vspace{3mm}
\caption{Rate regions for homogeneous and heterogeneous superposition coding.}
\label{table:comparison}
\end{table*}

\subsection{Heterogeneous Superposition Coding (UX Scheme)} 
\emph{Codebook generation:} 
Fix a pmf $p(u,x)$. Randomly and independently generate $2^{nR_1}$ sequences $u^n(m_1)$, $m_1 \in [1::2^{nR_1}]$, each according to $\prod_{i=1}^n p_U(u_i)$. For each message $m_1 \in [1::2^{nR_1}]$, randomly and conditionally independently generate $2^{nR_2}$ sequences $x^n(m_1,m_2)$, $m_2 \in [1::2^{nR_2}]$, each according to $\prod_{i=1}^n p_{X|U}(x_i|u_i(m_1))$.

\smallskip 
\emph{Encoding:} 
To send $(m_1,m_2)$, transmit $x^n(m_1,m_2)$. 

\smallskip 
\emph{Decoding:}
Both receivers use simultaneous nonunique decoding, which is rate-optimal as we show below.
In particular, upon receiving $y_1^n$, receiver~1 declares $\mh_1$ is sent if it is the unique message such that
$$
\big(u^n(\mh_1),x^n(\mh_1,m_2),y_1^n \big)
	\in\aep
$$
for some $m_2$.
If there is no unique $\mh_1$, it declares an error.  Similarly, upon receiving $y_2^n$,
receiver~2 declares $\mh_2$ is sent if it is the unique message such that
$$
\big(u^n(m_1),x^n(m_1,\mh_2),y_2^n \big)
	\in\aep
$$
for some $m_1$.
If there is no unique~$\mh_2$, it declares an error.  Standard
arguments show that receiver~1 will succeed if
\begin{align}
\label{eqn:ux1}
	R_1< I(U;Y_1) \qquad\text{or}\qquad
	R_1+R_2< I(X;Y_1),
\end{align}
or, equivalently, if
\begin{align*}
	R_1+\min\{R_2, I(X;Y_1 \cond U)\} < I(X;Y_1).
\end{align*}
Following an analogous argument to the one
in~\cite{Bandemer--El-Gamal--Kim2012a},  it can be shown that this
region cannot be improved by applying maximum likelihood decoding.

Receiver~2 will succeed if
\begin{align}
\label{eqn:ux2}
\begin{split}
R_2&\le I(X;Y_2\cond U)\\
R_1+R_2&\le I(X;Y_2).
\end{split}
\end{align}
In the Appendix, we show that this region cannot be improved by applying maximum likelihood decoding.
The regions for both receivers are depicted in Table~\ref{table:comparison}.  Let $\Rr_{UX}(p)$ denote the set of
all $(R_1,R_2)$ pairs satisfying both
\eqref{eqn:ux1}~and~\eqref{eqn:ux2}.  Clearly, the rate region
$$
\Rr_{UX}=\text{co}\bigg(\bigcup_{p\in\Pc_{UX}} \Rr_{UX}(p)\bigg)
$$
is achievable.  Here, $\Pc_{UX}$ is the set of distributions of the
form $p=p(u,x)$.  

If the roles of $m_1$ and $m_2$ in code generation are reversed, one
can also achieve the region $\Rr_{VX}=\text{co}(\cup_{p}\Rr_{VX}(p))$
obtained by swapping $Y_1$ with $Y_2$ and $R_1$ with $R_2$ in the
definition of $\Rr_{UX}(p)$.  

It is worth reiterating that the two schemes above differ only in the
dependence/independence between clouds around different $u^n$ sequences, and
not in the underlying distributions from which the clouds are
generated.  Indeed, it is well known that the classes of distributions
$\Pc_{UX}$ and $\Pc_{UV}$ are equivalent in the sense that for every
$p(u,x)\in\Pc_{UX}$, there exists a $q(u)\inpmfspace q(v)\inpmfspace q(x|u,v)\in\Pc_{UV}$ such
that $\sum_v q(u)\inpmfspace q(v)\inpmfspace q(x|u,v)=p(u,x)$ (see for 
example~\cite[p.~626]{El-Gamal--Kim2011}).

\vspace{3mm}
\section{Main Result} \label{sec:main}
\begin{theor}   \label{thm:inclusion}
The rate region achieved by homogeneous superposition coding includes the rate region achieved by heterogeneous superposition coding, i.e.,
	\begin{align*}
		\text{co}\big(\Rr_{UX}\cup\Rr_{VX}\big) &\subseteq \Rr_{UV}.
	\end{align*}
	Moreover, there are channels for which the inclusion is strict.  
\end{theor}

\begin{figure*}[!t]
\hspace{2em}
\subfigure[Rate region in~\eqref{eqn:union1}.]{
        \def\svgscale{1.0}
        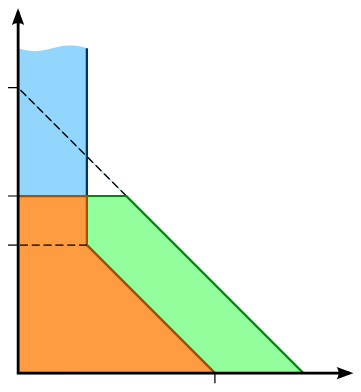
        \label{fig:case_ux}
}
\hspace{-5em}
\subfigure[Rate region in~\eqref{eqn:union2}.]{
        \def\svgscale{1.0}
        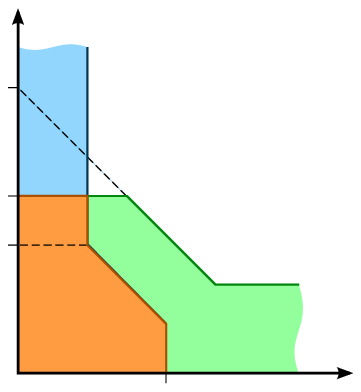
        \label{fig:case_uv}
}
\caption{Rate regions for the proof of Theorem~\ref{thm:inclusion}.}
\end{figure*}
\begin{proof}
Due to the convexity of $\Rr_{UV}$ and the symmetry between $UX$ and
$VX$ coding, it suffices to show that $\Rr_{UX}(p)\subseteq\Rr_{UV}$
for all $p\in\Pc_{UX}$.  Fix any $p\in\Pc_{UX}$. Let
$q'\in\Pc_{UV}$ be such that $U = X$, $V = \emptyset$, and the marginal on $X$ is preserved $q'(x)=p(x)$. Let
$q''\in\Pc_{UV}$ be such that $V = X, U = \emptyset$, and the marginal on $X$ is preserved $q''(x)=p(x)$.  An
inspection of \eqref{eqn:uv1}--\eqref{eqn:ux2} and
Table~\ref{table:comparison} reveals that $\Rr_{UV}(q')$ is the set
of rates satisfying
\begin{align*}
R_2&=0\\
R_1&\le I(X;Y_1),
\end{align*}
and $\Rr_{UV}(q'')$ is the set of rates satisfying 
\begin{align*}
R_1&=0\\
R_2&\le I(X;Y_2).
\end{align*}
It then follows that
$\text{co}\big(\Rr_{UV}(q')\cup\Rr_{UV}(q'')\big)$ includes the rate
region 
\begin{align}
\label{eqn:sum}
R_1+R_2\le \min\big\{I(X;Y_1),I(X;Y_2)\big\}.
\end{align}
We will consider three cases and show the claim for each.
\begin{itemize}
\item
If $I(X;Y_1)\ge I(X;Y_2)$  then $\Rr_{UX}(p)$ reduces to the rate
region
\begin{align*}
R_2&\le I(X;Y_2\cond U)\\
R_1+R_2&\le I(X;Y_2),
\end{align*}
which is included in the rate region in~\eqref{eqn:sum}, and therefore
in $\Rr_{UV}$.

\item
If $I(X;Y_1)< I(X;Y_2)$ and $I(X;Y_1\cond U)\ge I(X;Y_2\cond U)$, then
$\Rr_{UX}(p)$ reduces to the rate region
\begin{align*}
R_2&\le I(X;Y_2\cond U)\\
R_1+R_2&\le I(X;Y_1),
\end{align*}
which is also included in the rate region in~\eqref{eqn:sum}, and
therefore in $\Rr_{UV}$.

\item
If $I(X;Y_1)< I(X;Y_2)$ and $I(X;Y_1\cond U)<I(X;Y_2\cond U)$, then
$\Rr_{UX}(p)$ reduces to the rate region (see Figure~\ref{fig:case_ux})
\begin{align}
\label{eqn:union1}
\begin{split}
R_2 &\le I(X;Y_2\cond U)\\
R_1 + \min\{R_2,I(X;Y_1\cond U)\} &\le I(X;Y_1).
\end{split}
\end{align} 
Find a $q\in\Pc_{UV}$ with $q(u,x)=p(u,x)$, and note that
$\Rr_{UV}(q)$ is described by the bounds
\begin{align}
\label{eqn:union2}
\begin{split}
R_2&\le I(X;Y_2\cond U)\\
R_1&\le I(X;Y_1\cond V)\\
R_1+\min\{R_2,I(X;Y_1\cond U)\}&\le I(X;Y_1).
\end{split}
\end{align}
Comparing \eqref{eqn:union1}~with~\eqref{eqn:union2}
(Figure~\ref{fig:case_uv}), one sees that
$\Rr_{UX}(p)\subseteq\text{co}\big(\Rr_{UV}(q)\cup\Rr_{UV}(q')\big)$.
This proves the first claim of the theorem.
\end{itemize}
Now consider the vector broadcast channel with binary inputs
$(X_1,X_2)$ and outputs $(Y_1,Y_2)=(X_1,X_2)$.  For all
$p\in\Pc_{UX}$, we have from~\eqref{eqn:ux2} that $R_1+R_2\le
I(X_1X_2;Y_2)\le1$, and similarly for all $p\in\Pc_{VX}$.  Thus,
$
\big(\Rr_{UX}\cup\Rr_{VX}\big)$ is included in the rate
region $R_1+R_2\le1$.  Note, however, that the rate pair $(1,1)$ is
achievable using the $UV$ scheme by setting $U=X_1$ and $V=X_2$.  This
proves the second claim.  
\end{proof}

\vspace{3mm}
\section{Discussion}
\label{sec:remarks}

In addition to the basic superposition coding schemes presented in
Section~\ref{sec:superposition}, one can consider coded time
sharing~\cite{El-Gamal--Kim2011}, which could potentially enlarge the
achievable rate regions. In the present setting, however, it can be
easily checked that coded time sharing does not enlarge $\Rr_{UX}$.
Thus, the conclusion of Theorem~\ref{thm:inclusion} continues to hold
and homogeneous superposition coding with coded time sharing outperforms
heterogeneous superposition coding with coded time sharing.

\vspace{3mm}
\appendix[Optimality of the Rate Region in~\eqref{eqn:ux2}]  \label{sec:uxoptimality}
We show that no decoding rule for receiver 2 can achieve a larger rate region than the one in~\eqref{eqn:ux2} given the codebook ensemble of heterogeneous superposition coding. To this end, denote the random codebook by 
$$\Cc = (U^n(1),U^n(2),\dots, X^n(1,1), X^n(1,2), \dots). $$
By the averaged version of Fano's inequality in~\cite{Bandemer--El-Gamal--Kim2012a},
\begin{align}
	H(M_2 \cond Y_2^n, \Cc) &\leq n\eps_n,
	\label{eq:appdx_averageFanos}
\end{align}
where $\eps_n \to 0$ as $n\to\infty$. Thus,
{\allowdisplaybreaks
\begin{align*}
	nR_2 
	&= H(M_2) \\
	&\le I(M_2; Y_2^n \cond \Cc) + n\eps_n \\
	&\anneq{a} I(M_2; Y_2^n \cond \Cc, M_1) + n\eps_n \\
	&= H(Y_2^n\cond \Cc, M_1) - H(Y_2^n\cond \Cc, M_1, M_2) + n\eps_n \\
	&\annleq{b} nH(Y_2\cond U) - H(Y_2\cond X) + n\eps_n \\
	&= I(X; Y_2\cond U) + n\eps_n,
\end{align*}}%
where 
(a) follows by providing $M_1$ to receiver 2 as side information from a genie and
(b) follows from the codebook ensemble and the memoryless property.

To see the second inequality, first assume that 
\begin{align}
	R_1 < I(X;Y_2).
	\label{eq:appdx_R1small}
\end{align}
After receiver 2 has recovered $m_2$, the codebook given this message reduces to 
$$\Cc' = (X^n(1,m_2), X^n(2,m_2), X^n(3,m_2),\dots).$$
These codewords are pairwise independent since they do not share common $U^n$ sequences, and thus $\Cc'$ is a nonlayered random codebook. Since~\eqref{eq:appdx_R1small} holds, receiver 2 can reliably recover $M_1$ by using, for example, a typicality decoder. Thus,
by~\eqref{eq:appdx_averageFanos},
\begin{align*}
H(M_1,M_2\cond Y_2^n, \Cc) 
&= H(M_2 \cond Y_2^n, \Cc) + H(M_1 \cond Y_2^n, \Cc, M_2) \notag \\
&\le 2n\eps_n.
\end{align*}
Hence
\begin{align}
	n(R_1+R_2)
	&= H(M_1, M_2) \notag \\
	&\le I(M_1,M_2; Y_2^n \cond \Cc) + 2n\eps_n \notag \\
	&\le nI(X; Y_2) + 2n\eps_n.  \label{eq:appdx_R1R2bound}
\end{align}
To conclude the argument, assume there exists a decoding rule that achieves a rate point $(R_1,R_2)$ with $R_1 \geq I(X;Y_2)$. Then, this decoding rule must also achieve $(R'_1,R'_2) = (I(X;Y_2) - R_2/2, R_2)$, a rate point that is dominated by $(R_1,R_2)$. Since $R'_1 < I(X;Y_2)$, by our previous argument, $(R'_1,R'_2)$ must satisfy~\eqref{eq:appdx_R1R2bound}. It does not, which yields a contradiction.


\vspace{3mm}
\bibliographystyle{IEEEtranS}
\bibliography{nit}

\newcommand{\noopsort}[1]{}
\begin{thebibliography}{10}
\providecommand{\url}[1]{#1}
\csname url@samestyle\endcsname
\providecommand{\newblock}{\relax}
\providecommand{\bibinfo}[2]{#2}
\providecommand{\BIBentrySTDinterwordspacing}{\spaceskip=0pt\relax}
\providecommand{\BIBentryALTinterwordstretchfactor}{4}
\providecommand{\BIBentryALTinterwordspacing}{\spaceskip=\fontdimen2\font plus
\BIBentryALTinterwordstretchfactor\fontdimen3\font minus
  \fontdimen4\font\relax}
\providecommand{\BIBforeignlanguage}[2]{{%
\expandafter\ifx\csname l@#1\endcsname\relax
\typeout{** WARNING: IEEEtranS.bst: No hyphenation pattern has been}%
\typeout{** loaded for the language `#1'. Using the pattern for}%
\typeout{** the default language instead.}%
\else
\language=\csname l@#1\endcsname
\fi
#2}}
\providecommand{\BIBdecl}{\relax}
\BIBdecl

\bibitem{Bandemer--El-Gamal--Kim2012a}
B.~Bandemer, A.~El~Gamal, and Y.-H. Kim, ``Optimal achievable rates for
  interference networks with random codes,'' 2012, preprint available at
  http://arxiv.org/abs/1210.4596/.

\bibitem{Bergmans1973}
P.~P. Bergmans, ``Random coding theorem for broadcast channels with degraded
  components,'' \emph{{IEEE} Trans. Inf. Theory}, vol.~19, no.~2, pp. 197--207,
  1973.

\bibitem{Carleial1978}
A.~B. Carleial, ``Interference channels,'' \emph{{IEEE} Trans. Inf. Theory},
  vol.~24, no.~1, pp. 60--70, 1978.

\bibitem{Chia--El-Gamal2009}
Y.-K. Chia and A.~El~Gamal, ``3-receiver broadcast channels with common and
  confidential messages,'' in \emph{Proc. {IEEE} Int. Symp. Inf. Theory},
  Seoul, Korea, June/July 2009, pp. 1849--1853.

\bibitem{Chong--Motani--Garg--El-Gamal2008}
H.-F. Chong, M.~Motani, H.~K. Garg, and H.~El~Gamal, ``On the
  {H}an--{K}obayashi region for the interference channel,'' \emph{{IEEE} Trans.
  Inf. Theory}, vol.~54, no.~7, pp. 3188--3195, Jul. 2008.

\bibitem{Cover1972}
T.~M. Cover, ``Broadcast channels,'' \emph{{IEEE} Trans. Inf. Theory}, vol.~18,
  no.~1, pp. 2--14, Jan. 1972.

\bibitem{Cover--El-Gamal1979}
T.~M. Cover and A.~El~Gamal, ``Capacity theorems for the relay channel,''
  \emph{{IEEE} Trans. Inf. Theory}, vol.~25, no.~5, pp. 572--584, Sep. 1979.

\bibitem{Cover--Leung1981}
T.~M. Cover and C.~S.~K. Leung, ``An achievable rate region for the
  multiple-access channel with feedback,'' \emph{{IEEE} Trans. Inf. Theory},
  vol.~27, no.~3, pp. 292--298, 1981.

\bibitem{Csiszar--Korner1978}
I.~Csisz{\'a}r and J.~K{\"o}rner, ``Broadcast channels with confidential
  messages,'' \emph{{IEEE} Trans. Inf. Theory}, vol.~24, no.~3, pp. 339--348,
  1978.

\bibitem{El-Gamal--Kim2011}
A.~El~Gamal and Y.-H. Kim, \emph{Network Information Theory}.\hskip 1em plus
  0.5em minus 0.4em\relax Cambridge: Cambridge University Press, 2011.

\bibitem{Gallager1974}
R.~G. Gallager, ``Capacity and coding for degraded broadcast channels,''
  \emph{Probl. Inf. Transm.}, vol.~10, no.~3, pp. 3--14, 1974.

\bibitem{Grant--Rimoldi--Urbanke--Whiting2001}
A.~J. Grant, B.~Rimoldi, R.~L. Urbanke, and P.~A. Whiting, ``Rate-splitting
  multiple access for discrete memoryless channels,'' \emph{{IEEE} Trans. Inf.
  Theory}, vol.~47, no.~3, pp. 873--890, 2001.

\bibitem{Han--Kobayashi1981}
T.~S. Han and K.~Kobayashi, ``A new achievable rate region for the interference
  channel,'' \emph{{IEEE} Trans. Inf. Theory}, vol.~27, no.~1, pp. 49--60,
  1981.

\bibitem{Orlitsky--Roche2001}
A.~Orlitsky and J.~R. Roche, ``Coding for computing,'' \emph{{IEEE} Trans. Inf.
  Theory}, vol.~47, no.~3, pp. 903--917, 2001.

\bibitem{Ozarow1984}
L.~H. Ozarow, ``The capacity of the white {G}aussian multiple access channel
  with feedback,'' \emph{{IEEE} Trans. Inf. Theory}, vol.~30, no.~4, pp.
  623--629, 1984.

\end{thebibliography}

\end{document}